\documentstyle[prl,aps,epsfig]{revtex}         
\begin{document}
\draft               
\twocolumn[\hsize\textwidth\columnwidth\hsize\csname
@twocolumnfalse\endcsname
\title{Observation of Superfluid Flow in a Bose-Einstein Condensed Gas}
\author{R. Onofrio, C. Raman, J. M. Vogels, J. Abo-Shaeer, A. P. Chikkatur, and W. Ketterle}
\address{Department of Physics and Research Laboratory of
Electronics, \\
Massachusetts Institute of Technology, Cambridge, MA 02139}
\date{\today{}}
\maketitle

\begin{abstract}
We have studied the hydrodynamic flow in a Bose-Einstein
condensate stirred by a macroscopic object, a blue detuned laser
beam, using nondestructive {\em in situ} phase contrast imaging.
A critical velocity for the onset of a pressure gradient has been
observed, and shown to be density dependent. The technique has
been compared to a calorimetric method used previously to measure
the heating induced by the motion of the laser beam.
\end{abstract}

\pacs{PACS 03.75.Fi, 67.40.Vs, 67.57.De} \vskip1pc ]


Beginning with the London conjecture \cite{LONDON}, Bose-Einstein
condensation has been considered crucial for the understanding of
superfluidity. Since then, the weakly interacting Bose gas has
served as an idealized model for a superfluid \cite{STRINGARI}.
It has a phonon-like energy-momentum dispersion relation that
does not allow for the generation of elementary excitations below
a critical velocity, thus implying dissipationless flow at lower
velocities. The onset of dissipation has been treated in the
framework of the nonlinear Schr\"{o}dinger equation
\cite{FRISCH,HAKIM,HUEPE,WINIECKI,CARADOC} and shows an
intriguing richness. Experiments on liquid helium could not test
these theories, because superfluidity and dissipation are even
more complex in this system due to the presence of strong
interactions and surface effects \cite{WILKS}.

The creation of Bose-Einstein condensates in dilute gases has
dramatically changed this situation, allowing for quantitative
tests of microscopic theories using the tools and precision of
atomic physics experiments \cite{INGUSCIO}. A number of recent
experiments have examined phenomenological features of
superfluidity in gaseous Bose-Einstein condensates. These include
the observation of vortices \cite{MATTHEWS,MADISON}, a
non-classical moment of inertia \cite{MARAGO}, and suppression of
collisions from microscopic impurities \cite{CHIKKATUR}. In
previous work we found evidence for a critical velocity in a
stirred condensate \cite{RAMAN}. In this Letter we study the
onset of dissipation with higher sensitivity using repeated {\em
in situ} non-destructive imaging of the condensate. These images
show the distortion of the density distribution around the moving
object, thus directly probing the dynamics of the flow field that
has been recently treated with different models
\cite{JACKSON1,NORE,CRESCIMANNO,FEDICHEV,STIESBERGER,WINIECKIRECENT}.
The experimental setup was similar to the one used in our
previous work \cite{RAMAN}. Improvements in the evaporation
strategy and decompression techniques allowed us to produce pure
condensates with up to 5$\times10^7$ sodium atoms, with densities
ranging from 8.4$\times 10^{13}$ to 3.5$\times 10^{14}$
cm${}^{-3}$, corresponding to chemical potentials from 60 to 250
nK. We determined the Thomas-Fermi radius $R_z$ along the axial
direction through {\em in situ} phase contrast imaging. The sound
velocity at the center of the condensate was then evaluated
through the relationship $c_s=2 \pi \nu_z R_z/\sqrt{2}$, with
$\nu_z$=20.1 Hz being the axial trapping frequency. The
macroscopic moving object was a 514 nm laser beam blue-detuned
with respect to the sodium transitions, thereby creating an
effective repulsive potential for the atoms. The beam is focused
on the center of the condensate to a Gaussian $1/e^2$ diameter of
$2w=10\mu$m. It is scanned with an acousto-optic deflector driven
by a triangular wave along the axial direction of confinement of
the magnetic trap. This results in a motion of the laser beam at
constant speed $v=2af$, where $f$ is the frequency and $a$ is the
amplitude of the scan. To avoid effects due to the inhomogeneous
density distribution in the condensate, the laser beam was scanned
at various frequencies with a constant amplitude approximately
equal to the Thomas-Fermi radius $R_z$ of the condensate along
the axial direction.

\begin{figure}[htbf]
\epsfxsize=85mm \centerline{\epsfbox{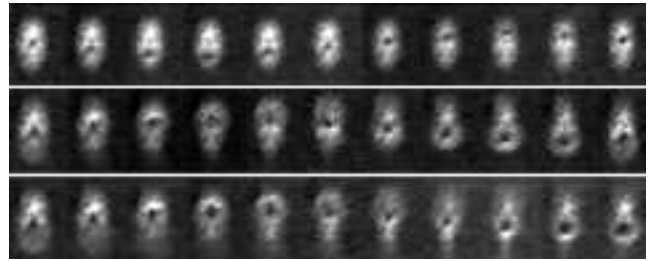}} \vspace{0.3cm}
\caption{Hydrodynamic flow in a Bose-Einstein condensate. Eleven
phase contrast images of a condensate stirred for one cycle are
taken {\em in situ} for various velocities of the laser beam. From
top to bottom: 0.4 mm/s, 7.0 mm/s, and 11.2 mm/s. The exposure
time is 300 $\mu$s per frame, and the frame rates are 33 Hz, 526
Hz, and 909 Hz, respectively. The maximum density and the
corresponding sound velocity without the laser are $2 \times
10^{14}$ cm${}^{-3}$ and 7 mm/s, respectively. The position of
the laser beam is seen as a depletion of the condensate around
its center. The vertical size of each image is $\simeq$ 200
$\mu$m.} \label{fig:movie_frames}
\end{figure}
\vspace{-0.3cm}

The frequency range was limited to 200 Hz. Above this frequency
the atoms experience an effective time-averaged stationary
potential. The condensate was imaged using the phase contrast
technique \cite{INGUSCIO,ANDREWS} with probe light red-detuned
from the F=1 to F${}^{\prime}$=2 transition. Different detunings
(0.4 and 1.7 GHz) allowed us to maintain a large linear dynamic
range when imaging condensates at different densities. Fig.
\ref{fig:movie_frames} shows multiple phase contrast images of
the hydrodynamic evolution of the density distribution of a single
condensate during the stirring. For low velocities there was
little effect on the condensate. However, at higher velocities
the density distribution became asymmetrical and the condensate
piled up in front of the moving beam, resulting in turbulent flow
around the object.

The main result of this Letter is based on quantitative analysis
of images like those in Fig.\ \ref{fig:movie_frames}. The
analysis of the observed flow pattern shows a critical velocity
for the onset of a drag force between the macroscopic laser beam
and the condensate.

\begin{figure}[htbf]
\epsfxsize=70mm \centerline{\epsfbox{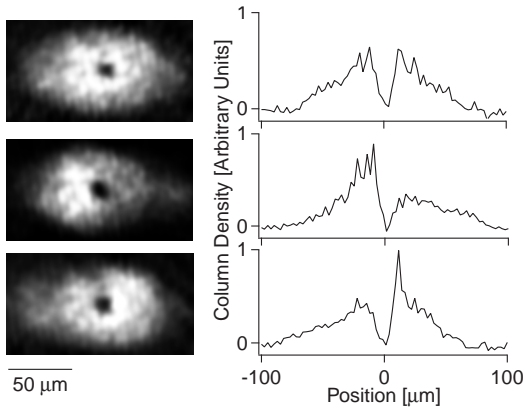}}\vspace{0.1cm}
 \caption{Pressure difference across a laser beam
moving through a condensate. On the left side {\em in situ} phase
contrast images of the condensate are shown, strobed at each
stirring half period: beam at rest (top); beam moving to the left
(middle) and to the right (bottom). The profiles on the right are
horizontal cuts through the center of the images. The stirring
velocity  and the maximum sound velocity were 3.0 mm/s and 6.5
mm/s respectively.} \label{fig:strobed}
\end{figure}
\vspace{-0.3cm}

For a weakly interacting Bose-condensed gas at density $n({\bf
r})$ and chemical potential $\mu$, pressure is identical to the
mean-field energy density $ P = \mu({\bf r}) n({\bf r})/2$
\cite{STRINGARI}. A drag force arises due to the pressure
difference across the moving object. The chemical potential is
given by $\mu({\bf r},t)=g n({\bf r},t)$, where $g = 4 \pi
\hbar^2 a /M$ is the strength of two-body interactions for atoms
with mass M and scattering length $a$. Therefore the magnitude of
the drag force $F$ is given by
\begin{equation} F \simeq  g S n {\bf \Delta} n=S \mu \Delta \mu/g
\label{eq:drag_force}
\end{equation}
where $\Delta n$ and $\Delta \mu$ are the differences in density
and chemical potential across the stirring object, and $S$ the
surface area the object presents to the condensate.

If the laser beam is stationary, or moves slowly enough to
preserve a possible superfluid state for the condensate, there
will be no gradient in chemical potential across the laser focus,
and therefore zero force according to Eq.(\ref{eq:drag_force}). A
drag force between the moving beam and the condensate is
indicated by an instantaneous density distribution $n(\bf{r},t)$
that is distorted asymmetrically with respect to the laser beam.
Fig.\ \ref{fig:strobed} shows this asymmetry as the formation of
a bow wave in front of and a stern wave behind the moving laser
beam.
\begin{figure}[btp]
\begin{center}\leavevmode
\epsfxsize=80mm \centerline{\epsfbox{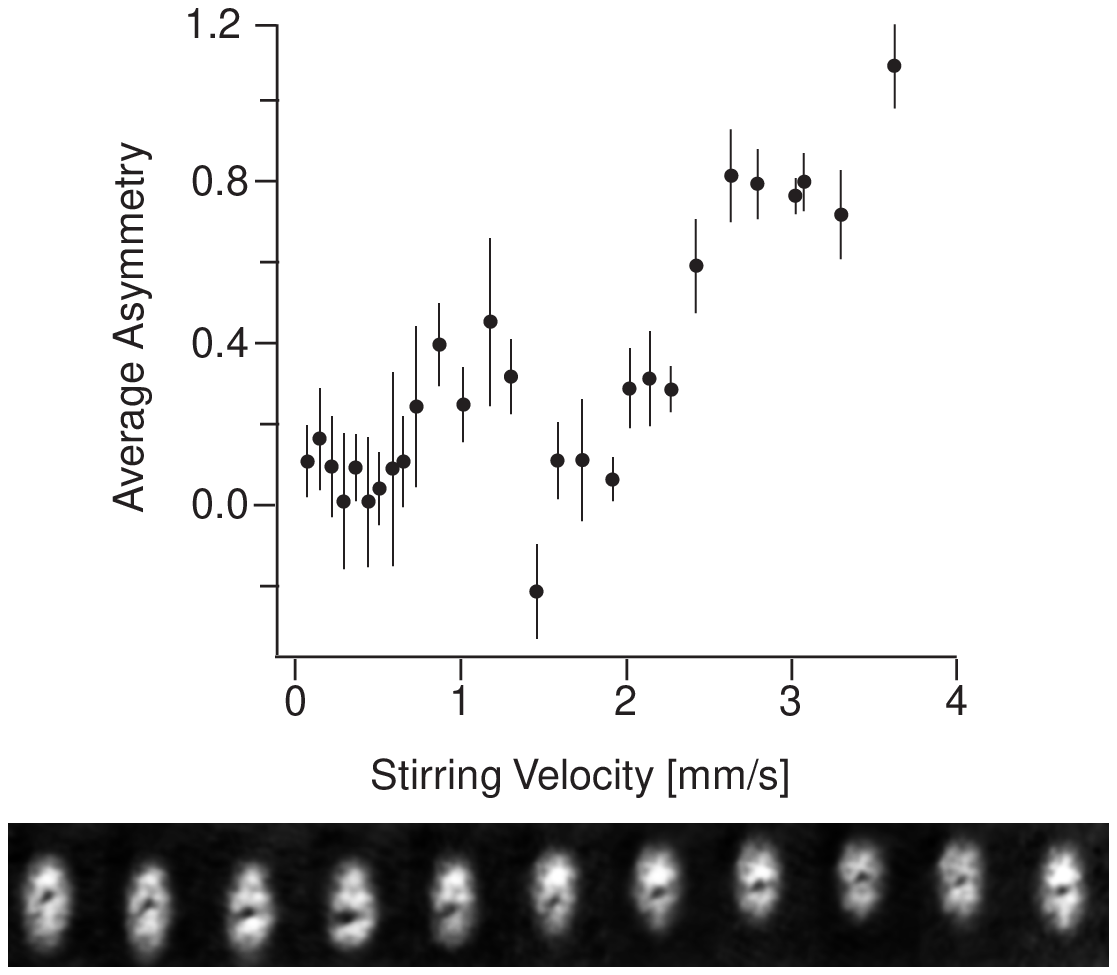}} \vspace{0.3cm}
\vspace{-10pt} \caption{Drag force on a Bose-Einstein condensate,
derived as an asymmetry in the density profile (top). Beyond 4
mm/s the signal saturated due to the formation of a high-density
bow wave, introducing nonlinearity into the phase contrast signal
for phase shifts exceeding $\pi$. The bars represent statistical
errors. Scanning the laser at the axial trapping frequency,
corresponding to a stirring velocity of 1.5 mm/s for an amplitude
of 37 $\mu$m, shows synchronous oscillations of the condensate
(bottom) and reduced heating.} \label{fig:asym_1}
\end{center}
\end{figure}
\vspace{-0.3cm}
 To disentangle asymmetries due to the moving beam from the
intrinsic inhomogeneity of the density profile we strobed the
images at half the stirring period and always viewed the laser
beam in the center of the condensate.  We defined the asymmetry
$A$ as the relative difference between the peak column densities
in front ($\tilde{n}^f$) and behind ($\tilde{n}^b$) the laser beam
$A = 2(\tilde{n}^{f}-\tilde{n}^{b})/(\tilde{n}^{f}+
\tilde{n}^{b})$. We averaged $A$ over 9 successive images of a
single condensate.

The average asymmetry versus velocity is shown in Fig.
\ref{fig:asym_1}. Below $\sim 0.6$ mm/s the asymmetry was zero
within the statistical error indicating a suppression of
dissipation in the superfluid regime. The asymmetry increased for
velocities greater than 0.6 mm/s.  The dip near 1.5 mm/s
corresponds to a stirring frequency of $\simeq$ 20 Hz for the
chosen amplitude. This is close to the axial frequency of the
magnetic trap. Near this frequency the laser beam synchronously
excited the axial dipole motion of the condensate (see Fig.
\ref{fig:asym_1}), thus reducing the {\em relative} velocity of
the stirrer and the condensate and therefore the drag. We
repeated the measurements at different stirring amplitudes to
disentangle velocity and frequency dependent effects. This
confirmed that the observed dip always occurred near 20 Hz,
whereas the other features of the observed asymmetries depended
only on velocity.

\begin{figure}[htbf]
\epsfxsize=70mm \centerline{\epsfbox{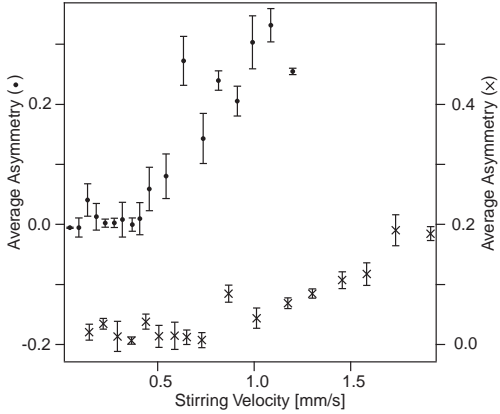}} \vspace{0.1cm}
\caption{Density dependence of the critical velocity. The onset
of the drag force is shown for two different condensate densities,
corresponding to maximum sound velocities of 4.8 mm/s ($\bullet$,
left axis) and 7.0 mm/s ({\bf $\times$}, right axis). The stirring
amplitudes are 29 $\mu$m and 58 $\mu$m, respectively.  The two
vertical axes are offset for clarity. The bars represent statistical errors.}\label{fig:asym_2}
\end{figure}
\vspace{-0.2cm}
 To study the superfluid regime at different atomic
densities, we focused on the low velocity region ($v \leq $ 2
mm/s) by choosing the stirring amplitudes in such a way that the
frequencies were below the trap resonance. The density was varied
by adjusting the number of atoms for constant trapping
parameters. The power in the laser beam was scaled to keep the
ratio of the potential height of the laser to the chemical
potential equal to 8 within 10\%, thus preserving the effective
size of the laser beam.

In Fig. \ref{fig:asym_2} we show measurements of the average
asymmetry for two maximum densities $n_0$ of $9 \times 10^{13}$
and $1.9 \times 10^{14}$ cm${}^{-3}$. Two features are evident.
In each data set there is a threshold velocity $v_c$ below which
the drag force is negligible, and this threshold increases at
higher density. Above this critical velocity, the drag force
increases monotonically, with a larger slope at low density.  The
data were fitted to a piecewise linear function that reflects the
two regimes of dissipation, one with zero asymmetry below $v_c$
and another, above $v_c$, where the asymmetry is linear in
$v-v_c$. The fits yield $v_c = (0.11\pm 0.02) c_s$ for the higher
density  and $v_c = (0.07\pm 0.02) c_s$ for the lower density.
This is in agreement with the expectation that the critical
velocity increases with the speed of sound and disagrees with
models that predict only a dependence upon  the diameter of the
object \cite{CRESCIMANNO,STIESBERGER}.

One can compare measurements of the asymmetry (proportional to the
drag force $\vec{F}$) with the power transferred to the condensate
by the moving beam, $\vec{F} \cdot \vec{v}$, using an improved
implementation of the calorimetric technique introduced in
\cite{RAMAN}. For this, we stirred the condensate for times
between 100 ms and 8 s, in order to produce approximately the
same final temperature. After the stirring beam was shut off, the
cloud was allowed to equilibrate for 100 ms. The thermal fraction
was determined using ballistic expansion and absorption imaging
\cite{INGUSCIO,RAMAN}. We inferred the temperature and total
energy using the specific heat evaluated as in \cite{GIORGINI}. A
further improvement was obtained by comparing successive images
with and without stirring of the condensate, to subtract out all
other sources of residual heating. As a result of the improved
fitting and background subtraction procedure, we were able to
detect changes of less than 10 nK in the energy of the gas, as we
will describe in more detail in \cite{RAMANJLTP}.  A condensed
fraction higher than 90\% ensured that the laser beam primarily
heated the condensate and not the thermal cloud.

\begin{figure}[htbf]
\epsfxsize=75mm \centerline{\epsfbox{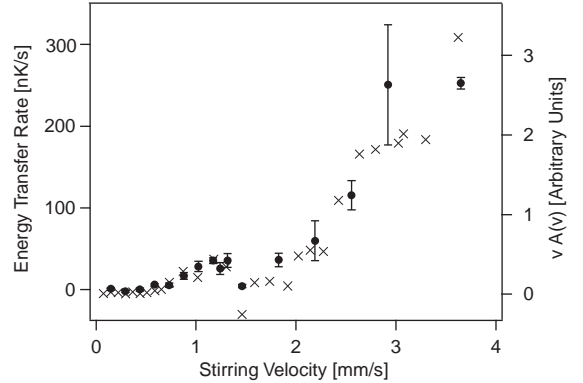}} \vspace{0.1cm}
\caption{Calorimetry of a condensate. The energy transfer rate
during stirring ($\bullet$, left axis) was obtained from
temperature measurements. The error bars reflect shot-to-shot
variations in the temperature. The results are compared to the
energy transfer rate $v A(v)$ obtained from the asymmetry data of
Fig. \ref{fig:asym_1} ($\times$, right axis).}
\label{fig:calorimetry}
\end{figure}

Fig. \ref{fig:calorimetry} shows the energy transfer rate to the
condensate versus the stirring velocity. Here, as in the
phase-contrast measurements, the amplitude of the stirring was
kept fixed while scanning the frequency.  We see the onset of
dissipation near the value obtained using the drag force method,
$v_c \simeq$ 0.5 mm/s = 0.1 $c_s$.

The calorimetric measurements can be compared with the drag force
inferred from the asymmetric density distribution. Using
Eq.(\ref{eq:drag_force}), the energy transfer rate per atom is
written in terms of the asymmetry as
\begin{equation}
{dE \over dt}= \frac{ \vec{F} \cdot \vec{v}}{N} \simeq
\frac{\mu_0 \tilde{n}_0 d}{N} v A(v) \label{eq:heatingasymmetry}
\end{equation}
Here we have used $S=  l_z d$ for the effective surface area of
the object, with $d$ the diameter of the laser beam, $l_z$ the
effective column depth, and $\tilde{n}_0=n_0 l_z$ for the observed
column density.

The comparison between the calorimetric  and the drag force measurements
 is also shown in Fig. \ref{fig:calorimetry}, with an adjusted
scale factor for $v A(v)$. The measurements of force (as an
asymmetry in the instantaneous condensate wavefunction) and
heating (as an integrated increase in thermal fraction) are in
remarkable agreement and demonstrates the consistency between our
two methods of analyzing dissipation. For the parameters of our
experiment ($d \simeq 10 \mu$m, $n_0=1.3 \times 10^{14}$
cm${}^{-3}$, $l_z =66 \mu$m, $N= 1.8 \cdot 10^{7}$) the overall
heating rate predicted by Eq.(\ref{eq:heatingasymmetry}) is 4.5
times larger than that obtained from calorimetry. This
discrepancy is not surprising, because
Eq.(\ref{eq:heatingasymmetry}) does not account for dynamical
pressure, the inhomogeneous density profile, and the 3D nature of
the flow field.

In one model \cite{FRISCH}, the drag force arises from the
periodic emission of vortex lines at a rate that increases with
velocity. The vortices reduce the pressure gradient across the
object and the predicted heating rate is \cite{FRISCH,WINIECKI}
\begin{equation}
{dE \over dt} \simeq \frac{1}{N} \frac{\mu}{h}
\frac{v}{c_s^2}(v-v_c) \epsilon_{vortex}
\end{equation}
where $\epsilon_{vortex}=2 \pi \tilde{n} \hbar^2/M \ln(d/\xi)$ is
the vortex energy  and $\xi = {(8 \pi n a)}^{-1/2}$ is the
healing length. Equating this to the heating rate in
Eq.(\ref{eq:heatingasymmetry}), predicts that the slope of the
asymmetry $\alpha = A(v)/(v-v_c)$ should scale as $\alpha \sim
1/c_s^2$. For the two densities in Fig. \ref{fig:asym_2}, the
model predicts the ratio $\alpha_{low}/\alpha_{high}$ = 2.1,
while our measurement yields $(1.9 \pm 0.2)$. This, and the
absolute value of the observed heating rate (see Ref.
\cite{RAMAN}), are further evidence that the drag force is closely
linked to vortex shedding.

Our new measurements are consistent with our earlier experiment
\cite{RAMAN} where dissipation at subsonic velocities was
observed.  However, the scatter in the observed heating rates in
\cite{RAMAN} prevented a clear observation of the threshold, and
the critical velocity was determined by linear extrapolation from
high heating rates to be $v_c/c_s = 0.25 \pm 0.04$.  Other
possible extrapolations tend to lower the value of $v_c$, for
instance a function $\propto v(v-v_c)$ as suggested by
Eq.(\ref{eq:heatingasymmetry}), yielded $v_c/c_s = 0.17 \pm 0.09$.
With improved calorimetry we have taken data for parameters
similar to \cite{RAMAN} in the range 0.1 $< v/c_s < $ 0.25  and
observed a small heating rate ($\simeq$ 20 nK/s at 1 mm/s) that
was previously indiscernible.

Since our first report on critical velocities in a Bose Einstein
condensate, several theoretical papers have added further insight
\cite{JACKSON1,NORE,CRESCIMANNO,FEDICHEV,STIESBERGER,WINIECKIRECENT}.
The observed critical Mach number $v_c/c_s \simeq 0.1$ is lower
than the predictions for homogeneous \cite{FRISCH,HUEPE} and
inhomogeneous \cite{WINIECKI} 2D systems. This discrepancy is
most likely due to the 3D geometry of our experiment, where the
laser beam pierced lower density regions of the condensate.  It
was noted in  Ref. \cite{FEDICHEV} that the critical velocity for
phonon excitation is lowered by the inhomogeneous density
distribution. Ref. \cite{NORE} shows that vortex stretching and
half-ring vortices can lower the 3D critical velocity below the
2D value. Jackson {\em et al.} \cite{JACKSON1} performed 3D
simulations for a geometry similar to our experiment and obtained
a critical velocity as low as $0.13 c_s$, in good agreement with
our results. The relevant critical velocity in our experiment is
most likely related to vortex {\it nucleation}
\cite{FRISCH,HUEPE,WINIECKI,CARADOC,JACKSON1,NORE,STIESBERGER,WINIECKIRECENT},
which is usually smaller than the Landau and Feynman critical
velocities \cite{WILKS} at which phonons \cite{FEDICHEV} or
vortices \cite{CRESCIMANNO} become energetically favorable.

In conclusion, we have demonstrated a new technique to study the
onset of dissipation in a stirred condensate. This method directly
measures density distributions and pressure differences. The
results obtained are consistent with calorimetric measurements
but are more immune to shot-to-shot noise because multiple
non-perturbative images of the flow field are taken. This
technique is well suited to shed further light on contradictory
predictions about the onset of dissipation in a gaseous superfluid
\cite{JACKSON1,NORE,CRESCIMANNO,FEDICHEV,STIESBERGER,WINIECKIRECENT}
and to study in detail how the drag force depends on condensate
size and density, as well as the object size.

We thank A. G\"orlitz, C. E. Kuklewicz and Z. Hadzibabic for
contributing to the early stages of this experiment, and D.
Kokorowski for a critical reading of the manuscript. This work
has been supported by ONR, NSF, ARO, NASA, and the David and
Lucile Packard Foundation. A.P.C. acknowledges additional support
from NSF. \vspace{-0.5cm}

\end{document}